\documentstyle[twocolumn,aps,prl,epsf,epsfig]{revtex}
\begin{document}
\draft
\title{Tripartite Entanglement
in a Bose Condensate by Stimulated Bragg Scattering}
\author{B. Deb and G.S. Agarwal}
\address{Physical Research Laboratory, Ahmedabad-380009, India}
\date{\today}
\maketitle
\begin{abstract}
We show that it is possible to entangle three different
many-particle  states by Bragg spectroscopy with nonclassical
light in a Bose condensate of weakly interacting atomic gases.
Among these three states, two are of atoms  corresponding to
two  opposite  momentum side-modes of the condensate;  and  the
other is of single-mode photons of the output probe beam. We
demonstrate strong dependence of the multiparticle entanglement
on the quantum statistics of the probe light. We
present detailed results on entanglement keeping in view of the
possible  experimental situation.
\end{abstract}

Entanglement is the most intriguing feature of quantum
mechanics. Of late, it has been recognized as a wonderful
resource for quantum information processing.  Creation of
various entangled states is the first  step towards development
of quantum communication.  In recent times, many schemes for
entenglement production  have been proposed and  demonstrated.
Entanglement in a Bose condensate arises quite naturally. For
instance, Bogoliubov's theory \cite{bogoliubov} of Bose-Einstein
conedensation (BEC) in weakly interacting gases  predicts that,
in the condensate ground state, a pair of particles moving with 
opposit momentum are absolutely correlated (entangled). This
nonclassical feature makes Bose condensates of weakly
interacting atomic gases \cite{bec} an excellent source  for
entanglement in massive particles. Many authors have proposed
the production of nonclassical states
\cite{sorensen,duan,pu,you,moore,raghavan,shi} in a
two-component Bose condensate. Bragg scattering has also been
considered \cite{gasenzer} for producing entanglement and
squeezing.

Entanglement between two particles is quite common, for example,
EPR states, polarization states of twin-photons, down converted
two-photon states  in optical parametric oscillator and so on.
In contrast, three particle entanglement is not so common,
though recently three-photon GHZ entangled states \cite{ghz} 
have been experimentally realized. Because of two dominent
momentum side-modes
$\mathbf{q}$ and $-\mathbf{q}$ involved in the Bragg
scattering,  a condensate seems to be a suitable candidate for
exploring tripartite entanglement among these two condensate
side-modes and the probe photon mode.  We demonstrate that Bragg
scattering with two input light beams - one pump and the other
probe, generates entanglement between the output probe and the
two  side-modes excited due to the
scattering of light. We analyze in detail the relevant parameter
regimes where such tripartite entanglement will show up. We
consider three different input probes: coherent, vacuum and
one-photon Fock state. Although vacuum field as an input has
been considered earlier and the entanglement between the
scattered field and one condensate side-mode ($\mathbf{q}$) is discussed
\cite{gasenzer}, here we report another interesting regime where
strong entanglement in long time limit persists.  Our results
show that although a vacuum or a coherent probe field can induce
entanglement between the scattered field and one excited
side-mode ($\mathbf{q}$) of the condensate, the other 
condensate side-mode (-$\mathbf{q}$)
remains immune to entanglement with the field. The most
interesting result we obtain is that a nonclassical one-photon
field state as an input probe can cause this other mode to
become entangled with the field and thus generates the desired
tripartite entanglement. We also find that nonclassical field is
useful for detecting the entanglement between the two condensate
side-modes.

There are mainly four  types of  processes in   pump-probe or
Bragg scattering of light by a condensate. First, one pump
photon is transformed into a probe photon causing conversion of
a zero-momentum atom  into an atom of  momentum $\mathbf{q}$. In
the second process, an atom moving with a  momentum
$-{\mathbf{q}}$  is scattered back into a zero-momentum atom by
the transformation of a  photon from pump  to probe mode. The
other two are the reverse processes of these two. Both the pump
and the probe  laser beams are detuned far off resonance from an
electronic excited state of the atoms in order to avoid heating
of the condensate. Because of interplay and simultainity of such
processes, Bragg spectroscopy   holds the key for generating
many-particle entanglement in the motional states of the atoms
and photons.

The Hamiltonian of the system $H=H_{A}+H_{F}+H_{AF}$, with
$H_{F} = \hbar\omega_{1}
\hat{c}_{\mathbf{k}_1}^{\dagger}\hat{c}_{\mathbf{k}_1} +
\hbar\omega_{2}
\hat{c}_{\mathbf{k}_2}^{\dagger}\hat{c}_{\mathbf{k}_2}$ and
\begin{eqnarray}
H_{A} &=& \sum_{k}\hbar\omega_{k}
\hat{a}_{{\mathbf{k}}}^{\dagger}\hat{a}_{{\mathbf{k}}}
\nonumber \\
&+& \frac{4\pi\hbar^2a_s}{2mV}
\sum_{\mathbf{k}_3,\mathbf{k}_4,\mathbf{k}_5,\mathbf{k}_6}
\hat{a}_{\mathbf{k}_3}^{\dagger}\hat{a}_{\mathbf{k}_4}^{\dagger}
\hat{a}_{\mathbf{k}_5}\hat{a}_{\mathbf{k}_6}
\delta_{\mathbf{k}_{3}+\mathbf{k}_4,\mathbf{k}_5+\mathbf{k}_6}
\\
\nonumber \\
 H_{AF} &=&\hbar\Omega
\hat{c}_{\mathbf{k}_2}^{\dagger}\hat{c}_{\mathbf{k}_1}
\sum_{k}\left(\hat{a}_{\mathbf{q+k}}^{\dagger}\hat{a}_{{\mathbf{k}}}
+\hat{a}_{-\mathbf{q}+\mathbf{k}}\hat{a}_{{\mathbf{k}}}^{\dagger}
\right) +
{\rm H.c.} \label{hI}
\end{eqnarray}
where $\hat{c}_{\mathbf{k}} (\hat{c}_{\mathbf{k}}^{\dagger})$
represents the annihilation(creation) operator   for a laser
photon with momentum  $\mathbf{k}$,
$\hat{a}_{k}(\hat{a}_{k}^{\dagger})$ is the annhilation(creation)
operator for atoms with momentum $\mathbf{k}$ and  frequency
$\omega_{k}=\frac{\hbar k^2}{2m}$;
$\Omega=(\vec{E}_{1}.\vec{d}_{13})
(\vec{E}_{2}.\vec{d}_{32})/\Delta$ is the two-photon Rabi
frequency, where $E_{1(2)}$ is the  the pump(probe) field
amplitude with frequency $\omega_{1(2)}$ and  the $\vec{d}_{ij}$
is the electronic transition dipole moment between the states
$|i\rangle$ and $|j\rangle$ of an atom. Here $a_s$ is the s-wave
scattering length of the atoms and $V$ is the volume of the
condensate.

We assume that zero-momentum ($\mathbf{k}=0$) condensate state is
macroscopically occupied and therefore the atom-atom interaction
characterized by s-wave scattering length for a weakly interacting
atomic gas  is mainly due to the collision between zero- and and
non-zero momentum atoms. Applying Bogoliubov's prescription
$\hat{a}_{0},\hat{a}_{0}^{\dagger} \rightarrow \sqrt{N_{0}}$,
while the condensate fraction $N_{0}/N$, with $N$ being the total
number of atoms, and the number density $n_{0}=N_{0}/V$ remaining
fixed in the thermodynamic limit, we  convert the hamiltonian
$H_{A}$ into a quadratic form \cite{fetter}.
Further, we apply Bogoliubov's transformation \cite{fetter} between
particle and quasi-particle operators: $\hat{a}_{k}=
u_{k}\hat{\alpha}_{k}-v_{k}\hat{\alpha}_{-k}^{\dagger}$ where
$v_{k} = (u_{k}^{2}-1)^{1/2}
= [\frac{1}{2}(\frac{\omega_{k} + \mu/\hbar}{\omega_{k}^B}-1)]^{1/2}$
and
\begin{eqnarray}
\omega_{k}^B=\left[(\omega_{k} + \frac{\mu}{\hbar})^2 -
(\frac{\mu}{\hbar})^2\right]^{1/2},
\end{eqnarray}
is energy of Bogoliubov's  quasi-particle. Here
$\mu = \frac{\hbar^2\xi^{-2}}{2m}$ is the chemical potential
with
$\xi = (8\pi n_{0}a_{s})^{-1/2}$ being a characteristic
length scale known as the healing length. The condensate
ground state energy is
$E_g=\frac{1}{2}N_{0}\mu - \frac{1}{2}\sum_{k\ne
0} v_k^2E_k$.
We thus diagonalize $H_{A}$ and rewrite the entire Hamiltonian in
terms of Bogoliubov's  quasi-particle operators
$\hat{\alpha}_{\mathbf{k}}$. We  consider the condensate ground state
energy as the zero of the energy scale. By treating the pump light
beam classically,  the effective Hamiltonian can be written in
the standard form
\begin{eqnarray}
H_{eff} = &\hbar
\omega_{q}^{B}&\left(\hat{\alpha}_{{\mathbf{q}}}^{\dagger}
\hat{\alpha}_{{\mathbf{q}}}
+\hat{\alpha}_{-{\mathbf{q}}}^{\dagger}
\hat{\alpha}_{-{\mathbf{q}}}\right)
-\hbar\delta\hat{c}_{\mathbf{k}_2}^{\dagger}
\hat{c}_{\mathbf{k}_2} \nonumber \\
&+&\left[\hbar\eta
\hat{c}_{\mathbf{k}_{2}}^{\dagger}
(\hat{\alpha}_{\mathbf{q}}^{\dagger} +
\hat{\alpha}_{-{\mathbf{q}}})
+{\mathrm H.c.}\right]
\end{eqnarray}
where $\delta=\omega_{1}-\omega_{2}$, ${\mathbf{q}} =
{\mathbf{k}}_{1} - {\mathbf{k}}_{2}$ and $\eta =
\sqrt{N}f_q\Omega$; where  $f_q = u_q-v_q$. In writing Eq.(4),
we have retained only two dominent momentum side-modes of the
condensate, and neglected all other modes under the Bragg
resonance condition ($\delta \simeq \omega_{q}$). Equation (4)
involves three coupled operators. The evolution operator
$\exp(-iH_{eff}t/\hbar)$ can not be disentangled into those of
individual operators, since there exists no such disentanglement
theorem for such Hamiltonian. We therefore, emphasize that the Hamiltonian
(4) can be solved exactly and more easily in Heisenberg picture.
The Heisenberg equations of motion for a triad of operators $X =
\left(\hat{\alpha}_{{\mathbf{q}}} \hspace{0.2cm}
\hat{\alpha}_{-{\mathbf{q}}}^{\dagger} \hspace{0.2cm}
\hat{c}_{{\mathbf{k}}_{2}}^{\dagger}\right)^{T}$ can be written
in a matrix form $\dot{X} = i\omega_{q}^{B}{\mathbf{M}}X$. The
dynamics of these three coupled modes (two momentum side modes
$\pm \mathbf{q}$ plus one probe field mode) is controlled by the
eigenvalues of $\mathbf{M}$ matrix. The long time behavior is
oscillatory or growing depending on the momentum transfer $q$
and the effective coupling constant $\eta$. For a given value of
$q$, if $\eta$ exceeds a threshold value $\eta_{th}$, the
long-time dynamics becomes hyperbolic dominated by the complex
eigenvalues. In contrast, if $\eta$ is less than $\eta_{th}$, the
long time behavior  is oscillatory. Fig.1 shows the dependence
of $\eta_{th}$ on the momentum $q$ of the recoiling atoms. In
the phonon or quasi-particle regime ($\xi q<1$), $\eta_{th}$ is
much larger compared to that in particle regime ($\xi q >\!>1$).
For experimental situation of Ref.\cite{phonon} with  Na
condensate, where quasi-particle regime ($\xi q \simeq 0.47$) is
probed by Bragg scattering of light, $\eta_{th} = 0.314 
\omega_{q}^{B}$ with $\omega_{q}^{B} = (2\pi)\times 4.7$ kHz.
The value of $\eta$ used in that experiment is about 150
$\omega_{q}^{B}$.
\begin{figure}
\psfig{file=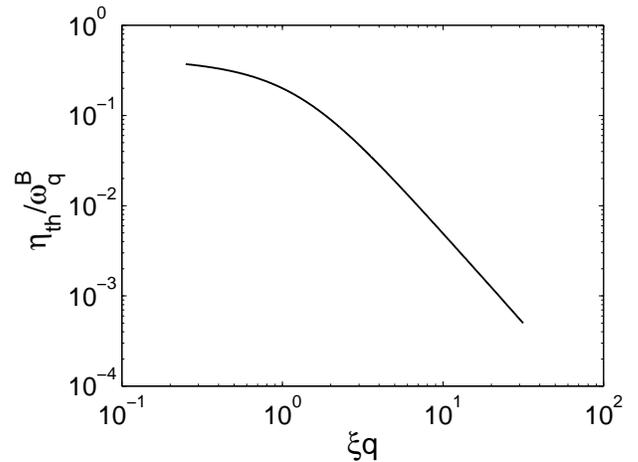,width=3.25in} 
\caption{The threshold value of
$\eta$ ($\eta_{th}/\omega_{q}^{B}$) as a function of $\xi q$.}
\end{figure}

To demonstrate many-particle entanglement among two motional
(momentum) states of the condensate and one probe field state,
following Ref.\cite{gasenzer,wineland}, we define two-mode
entanglement parameter
\begin{equation}
\xi_{i,j} = \langle[\Delta(\hat{n}_{i}-\hat{n}_{j})]^2\rangle/
(\langle\hat{n}_{i}\rangle + \langle\hat{n}_{j}\rangle), \hspace{0.2cm}
i,j=q,-q,k_{2}
\end{equation}
where $\langle (\Delta \hat{n})^2 \rangle = 
\langle\hat{n}^2\rangle - \langle\hat{n}\rangle^{2}$ and
$\hat{n}_{i,j}$ are the two particle number operators
$\hat{a}_{\mathbf{q}}^{\dagger}\hat{a}_{\mathbf{q}},
\hat{a}_{-\mathbf{q}}^{\dagger}\hat{a}_{-\mathbf{q}}$ and one
photon number operator
$\hat{c}_{\mathbf{k}_{2}}^{\dagger}\hat{c}_{\mathbf{k}_{2}}$.
The particle operators $\hat{a}$ are related to the
quasi-particle operators $\hat{\alpha}$ by Bogoliubov's
transformation. If $\xi_{i,j}$ is below unity, the corresponding
two-modes 'i', and 'j' are entangled.  We specifically consider a
condensate of $5\times 10^{6}$ Na atoms. First, we examine
entanglement by light scattering in phonon regime ($\xi q < 1$).
While choosing  the relevant parameters, we followed  the
experiment of Ref.\cite{phonon}, i.e.,   chemical potential $\mu
= 6.7$ kHz,  the pump and the probe fields intersecting at an
angle $14^{0}$ transfer a momentum  $q = 0.47 \xi^{-1}$ to the
atoms. Figure 2 illustrates the  different two-mode entanglement
parameters as a function of time with probe field being
initially in a {\bf coherent state} $|\beta\rangle $ with
average photon number $|\beta|^2 = 1.0$. Compared to long-time
hyperbolic limit (Fig.2(a)), the oscillatory limit (Fig.2(b))
seems to be more interesting from the two-mode or multimode
entanglement viewpoint. As shown by Fig.2(b), entanglement
between the ($q, -q$) and ($q, k_{2}$) revives and disappears
with time in the long-time limit, while in the hyperbolic limit,
entanglement completely vanishes after a certain time
(Fig.2(a)). Before vanishing, $\xi_{q,k_{2}}$ makes small
oscillations eventually reaching a minimum (maximum
entanglement). As the coupling $\eta$ increases, these
oscillations become less prominent (inset to Fig.2(a)), while for
higher momentum ($\xi q >1$), these oscillations
completely die down (inset to Fig.2(b)). The oscillations are
caused by the interference of the three closely lying 
eigenvalues of $\mathbf{M}$. From Fig.2, we infer that although the two
condensate  side-modes $\mathbf{q}$ and -$\mathbf{q}$ and one side-mode  $q$ and
the scattered field mode ${\mathbf{k}}_{2}$ are entangled in different time
and parameter zones, modes -$\mathbf{q}$ and 
${\mathbf{k}}_{2}$ does not exhibit any
mutual entanglement in any zone. However, by changing the
character of the applied probe field, as we show below, it is
possible to obtain mutual entanglement among all the three
modes. By comparing Fig.2(b) with Fig.3(a) which exhibits
$\xi_{ij}$ when the probe field is in {\bf vacuum}, we further
infer that Bragg scattering with either coherent or vacuum probe
field generates almost similar entanglement characteristics in
the system. Although there arises mutual entanglement between
the modes ($\mathbf{q}, -\mathbf{q}$) and ($\mathbf{q},
{\mathbf{k}}_{2}$) due to either of theses two fields,
entanglement between  ($-\mathbf{q}, {\mathbf{k}}_{2}$) is not
developed in either case. The results depicted in Fig.3(a) may
be contrasted with those of Ref.\cite{gasenzer} which studied
entanglement between $\mathbf{q}$ and ${\mathbf{k}}_{2}$ modes
in  the hyperbolic limit for a vacuum probe. In contrast, the
oscillatory limit we describe in the present paper is a new
result not discussed before thus far. We emphasize that
entanglement in the oscillatory limit is more significant; since
it sustains at long time, while in the hyperbolic limit it does
not exist at long time. We also point out that these two  limits
depend on the two parameters $q$ and $\eta$, and not due  the
dynamical processes of the three coupled modes.

Next, we consider the probe field as a {\bf one-photon Fock
 state} \cite{onephoton}. In this case, all the three modes
 exhibit mutual entanglement at different times and duration as
 is evident from the Fig.3(b). Thus three-mode or tripartite 
 entanglement can be generated  in a Bose condensate by Bragg
 spectroscopy with nonclassical light fields (Fock states) which
 constitutes another central result of the present paper. It
 should be pointed out that neither vacuum nor coherent probes
 are able to generate such tripartite entanglement in any
 parameter space. Why one-photon field state and neither vacuum
 nor the coherent state can generate tripartite entanglement
 may be explained by examining the respective phae-space density
 distribution functions. Since, coherent state is a displaced
 vacuum state, both of them have almost similar phase-space
 structure, except a displacement of the equilibrium position.
 We have  developed a formalism based on time-dependent Wigner
 distribution functions  which provides significant insight into
 the entanglement properties in various cases discussed in this
 paper.
\begin{figure}
\psfig{file=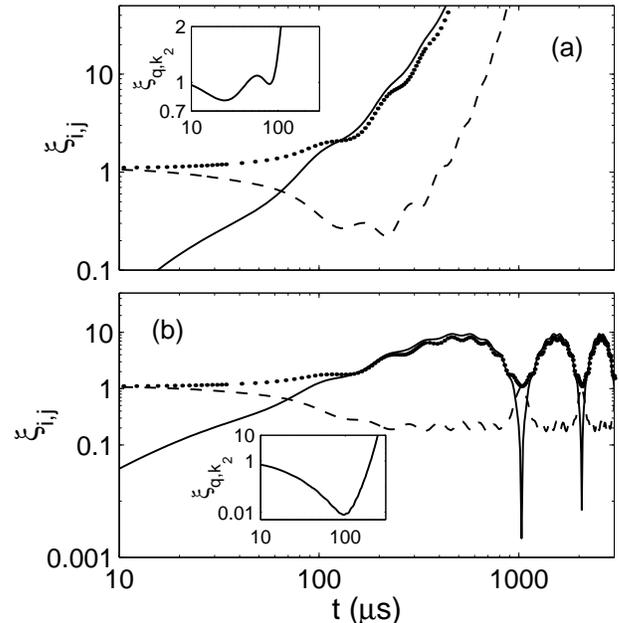,width=3.25in}
 \caption{(a) Two-mode entanglement parameter $\xi_{q,-q}$
(solid line),   $\xi_{q,k_{2}}$ (dashed lines) and
$\xi_{-q,k_{2}}$ (dotted line) as a function of time $t$ in
microsecond ($\mu s$). Initially, the condensate is assumed to
be  in the ground state  and the probe field is in a coherent
state with average photon number $\langle n_{p}\rangle = 1$. The
other parameters: $ q = 0.47 \xi^{-1}$,  $\omega_{q}^{B} =
(2\pi)\times 4.7$ kHz, the two-photon Rabi frequency $\Omega =
8$Hz,  $\eta/\omega_{q}^{B}=0.34$ corresponding to hyperbolic
regime, since  $\eta_{th}/\omega_{q}^{B} = 0.313$. The
eigenvalues of $\mathbf{M}$ matrix  are 1.07975, - 0.69788 +
$0.14153i$ and  -0.69788 - $0.14153i$.  (b) Same as in (a), but
$\Omega = 7 $ Hz corresponding to oscillatory regime
($\eta/\omega_{q}^{B} = 0.29$) with eigenvalues 1.06235,
-0.79248, -0.58587.  The  inset to Fig.(a) shows the variation of
$\xi_{q,k_{2}}$ with $t$ in $\mu s$ for $\Omega = 16$ Hz and $ q
= 0.47 \xi^{-1}$.  The inset to (b): $\xi_{q,k_{2}}$ Vs $t$ in
$\mu s$ for $\Omega = 16$ Hz, $q = 2 \xi^{-1}$.} \label{fig2}
\end{figure}

Now, to know the entanglement property in the particle regime
 ($\xi q>>1$), we show the $\xi_{ij}$ as a function of time for
 $q = 8.329\xi^{-1}$ in Fig.4. While considering phonon
 ($\omega_{q}^{B} \propto q$) or particle ( $\omega_{q}^{B}
 \propto q^2)$ regime, exact value of $\omega_{q}^{B}$ should be
 taken, otherwise an approximation in $\omega_{q}^{B}$ may lead
 to wrong results if $\eta$ is close to the threshold value
 $\eta_{th}$. Here also we consider a Na condensate of $ 5\times
 10^{6}$ atoms, but the other parameters ($\mu, q$) are
 chosen from the experimental paper of Ref.\cite{inouye} From
 Fig.4(a), we observe that before reaching the hyperbolic
 regime, all the three modes remain entengled for a considerable
 duration. The inset to Fig.4(a) shows the variation of
 $\xi_{ij}$ as a function of the two-photon Rabi frequency
 $\Omega$ at time $t = 10 \mu$s. From this figure we conclude
 that the tripartite entanglement is signicant at weak coupling
 (low $\Omega$ or low $\eta$) near the threshold ($\eta_{th}$).
 Figure 4(b) exhibits entanglement property in the same particle
 regime but in the long-time oscillatory limit. In this case,
 unlike the other cases, entanglement among the three modes
 persists for almost all the time, although in the long-time limit  the
 modes $\mathbf{q}$ and $-\mathbf{q}$ (solid curve) are
 marginally entangled. But, in contrast, the modes $\mathbf{q}$
 and ${\mathbf{k}}_{2}$ remain largly entangled for all the
 time.
\begin{figure}
\psfig{file=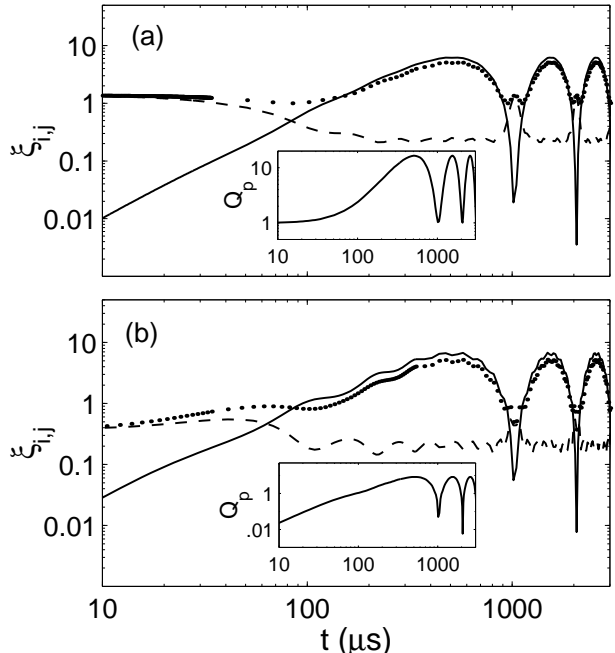,width=3.25in}
 \caption{ (a) Same as in Fig.2(b), but for the probe field being
 initially in vacuum (cavity). The inset to Fig.3 (a) shows the
 $Q$ parameter $Q_{p}$ of the output probe field as a function of
 time in $\mu s$. This $Q$ parameter always remains greater than
 unity. (b) Same as in (a), but for probe field being
 initially in one photon Fock state. The inset to Fig.3 (b) shows
 the corresponding $Q_{p}$ as a function of time in $\mu s$. In
 this case, the $Q$ parameter becomes less than unity implying
 nonclassicality of the output light field.}
\label{fig3}
\end{figure}
To see whether the photon number fluctuation of the output
 (scattered) light field has any connection with the
 entanglement parameters discussed so far, we calculate the
 Mandel $Q$ parameter $Q_{p} = \langle (\Delta
 \hat{n}_{{\mathbf{k}}_{2}})^2 \rangle /\langle
 \hat{n}_{{\mathbf{k}}_{2}} \rangle$ We find that neither a
 coherent nor a vacuum field as an input probe can generate
 nonclassical light output, while a nonclassical one-photon Fock
 state can do so. The time-evolution of the $Q$ parameter has a
 strong link with the time-evolution of the entanglement
 parameter $\xi_{q,-q}$, i.e., entanglement between the atomic
 modes $\mathbf{q}$ and $-\mathbf{q}$. With a nonclassical input
 probe, both   the parameters $Q_{p}$ and $\xi_{q,-q}$ become
 less than unity at the same time (Fig.3 (b) inset) indicating
 that the entanglement between the two atomic side-modes may be
 inferred by measuring $Q_{p}$  using beam splitter at the
 output of the probe beam and detecting the photon number
 fluctuations with two detectors.

In conclusion, we have presented an exact treatment of the
dynamics of excitations in a condensate  by Bragg scattering of
light under Bragg resonance condition. We have shown that by
using a nonclassical light (Fock states) as input probe, it is
possible to generate tripartite entanglement in the
many-particle states of three motional modes which include two
momentum side-modes of the condensate and one probe field mode.
The relevant physical parameter regimes where such tripartite
entanglement can be observed are analyzed. The important
question which needs  detailed study  is how to detect the
generated entanglement. Our results suggest that by measuring
the second order correlation function of the scattered field, it
is possible to measure the entanglement between the two modes of
the condensate. The number variance of the $\mathbf{q}$ and
$-\mathbf{q}$ momentum atoms may also be measured by outcoupling
them with a change in the hyperfine spin state by applying
Doppler-sensitive  two radio frequency pulses and then by
measuring the intensity variation at the output of the two
pulses.
\begin{figure}
\psfig{file=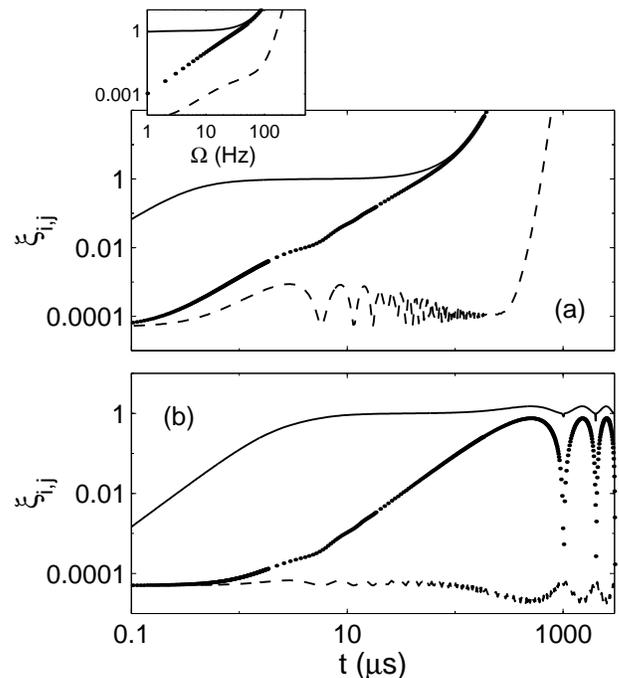,width=3.25in}
 \caption{(a) Same as in Fig.3(b), but $\mu = (2\pi)\times 1.23$ kHz,
 $q = 8.329 \xi^{-1}$, $\omega_{q}^{B} = (2\pi)\times 86.65 
 k$Hz, $\Omega = 7$ Hz, $\eta =
 0.0285 \omega_{q}^{B}$ and $\eta_{th} =0.0071 \omega_{q}^{B}$.
 The eigenvalues are 1.00041, -0.99315 + $0.02771i$, -0.99315 - $0.02771i$.
  (b) Same as in (a), but $\Omega = 1 $ Hz corresponding to
 oscillatory regime ($\eta = 0.0041 \omega_{q}^{B}$) with eigenvalues
 1.00001, -0.9987, -0.9872.
 The  inset to Fig.(a) shows the variation of
 $\xi_{ij}$ with two-photon Rabi frequency  $\Omega $ in Hz at
 time $t = 10 \mu$s with other
 parameters remaining the same as in the main figure of (a).}
\label{fig1}
\end{figure}

\end{document}